\documentclass[12pt,leqno]{amsart}
\usepackage{amssymb}
\usepackage{amsmath,amssymb}
\newtheorem{theorem}{Theorem}[section]
\newtheorem{corollary}{Corollary}[section]

\begin{document}
\theoremstyle{plain}
\newtheorem{MainThm}{Theorem}
\newtheorem{thm}{Theorem}[section]
\newtheorem{clry}[thm]{Corollary}
\newtheorem{prop}[thm]{Proposition}
\newtheorem{lem}[thm]{Lemma}
\newtheorem{deft}[thm]{Definition}
\newtheorem{hyp}{Assumption}
\newtheorem*{ThmLeU}{Theorem (J.~Lee, G.~Uhlmann)}

\theoremstyle{definition}
\newtheorem{rem}[thm]{Remark}
\newtheorem*{acknow}{Acknowledgements}
\numberwithin{equation}{section}
\renewcommand{\d}{\partial}
\newcommand{\re}{\mathop{\rm Re} }
\newcommand{\im}{\mathop{\rm Im}}
\newcommand{\R}{\mathbf{R}}
\newcommand{\C}{\mathbf{C}}
\newcommand{\N}{\mathbf{N}}
\newcommand{\D}{C^{\infty}_0}
\renewcommand{\O}{\mathcal{O}}
\newcommand{\dbar}{\overline{\d}}
\newcommand{\supp}{\mathop{\rm supp}}
\newcommand{\abs}[1]{\lvert #1 \rvert}
\newcommand{\csubset}{\Subset}
\newcommand{\detg}{\lvert g \rvert}
\title[Reconstruction of Lam\'e coefficients]
 {On reconstruction of Lam\'e coefficients from partial Cauchy
data in three dimensions}

\author[O. Imanuvilov]{Oleg Yu. Imanuvilov}
\address{Department of Mathematics, Colorado State
University, 101 Weber Building, Fort Collins
CO, 80523 USA\\
e-mail: oleg@math.colostate.edu}
\thanks{First author partly supported
by CGO program of Univ at Tokyo}

\author[G. Uhlmann]{Gunther Uhlmann}
\address{Department  mathematics, UC Irvine, Irvine CA 92697\\
Department of Mathematics, University of Washington, Seattle,
WA 98195 USA\\
e-mail: gunther@math.washington.edu}
\thanks{Second author partly supported by
NSF}

\author[M. Yamamoto]{Masahiro Yamamoto}
\address{Department of Mathematical Sciences,
The University of Tokyo, Komaba, Meguro, Tokyo 153, Japan \\e-mail:
myama@ms.u-tokyo.ac.jp}

\begin{abstract}
For the isotropic Lam\'e system, we prove in dimensions three or larger
that both Lam\'e coefficients are uniquely recovered from partial Cauchy data
on an arbitrary open subset of the boundary provided that the coefficient $\mu$
is a priori close to a constant.
\end{abstract}

\maketitle \setcounter{tocdepth}{1} \setcounter{secnumdepth}{2}

In a bounded domain $\Omega \subset {\R}^N, N\ge 3$ with smooth
boundary we consider the isotropic Lam\'e system:
\begin{equation}\label{lam0}
\sum_{j,k,l=1}^N\frac{\partial}{\partial
x_j}\left(C_{ijkl}\frac{\partial u_k}{\partial x_l}\right)=0\quad
\mbox{in}\,\,\Omega, \thinspace 1\le i\le N
\end{equation}
and
\begin{equation}\label{lam}
u\vert_{\partial\Omega}=f,
\end{equation}
where
$$
C_{ijkl}=\lambda(x)\delta_{ij}\delta_{kl}+\mu(x)(\delta_{ik}\delta_{jl}
+\delta_{il}\delta_{jk}), \quad 1\le i,j,k,l\le N
$$
where the Kr\"onecker delta is denoted by $\delta_{ij}$. The
functions $\lambda$ and $\mu$ are called the Lam{\' e} coefficients,
$u(x)=(u_1(x), \dots, u_N(x))$ is the displacement.  Assume that
\begin{equation}\label{Lame}
\mu(x)>0\quad \mbox{on}\,\,\overline\Omega, \quad
(\lambda+\mu)(x)>0\quad \mbox{on}\,\,\overline\Omega .
\end{equation}

We set
$$
\Lambda_{\lambda,\mu}u =
\left(\sum_{j,k,l=1}^N\nu_jC_{1jkl}\frac{\partial u_k}{\partial
x_l},\dots, \sum_{j,k,l=1}^N\nu_jC_{Njkl}\frac{\partial
u_k}{\partial x_l}\right),
$$
where $\nu=(\nu_1,\dots, \nu_N)$ is the outward unit normal vector
to $\partial \Omega.$ The map $\Lambda_{\lambda,\mu}u$ is called the
Dirichlet-to-Neumann map that maps the displacement at the boundary
to the traction. Denote
$$
\mathcal L_{\lambda,\mu}(x,D) u =
\left(\sum_{j,k,l=1}^N\frac{\partial}{\partial
x_j}\left(C_{1jkl}\frac{\partial u_k}{\partial x_l}\right),
\dots,\sum_{j,k,l=1}^N\frac{\partial}{\partial
x_j}\left(C_{Njkl}\frac{\partial u_k}{\partial x_l}\right)\right).
$$
The partial Cauchy data $\mathcal C_{\lambda,\mu}$ is defined by
$$
\mathcal C_{\lambda,\mu}=\{ (u,\Lambda_{\lambda,\mu}
u)\vert_{\widetilde \Gamma}; \thinspace \mathcal
L_{\lambda,\mu}(x,D) u=0\quad \mbox{in}\,\,\Omega,\,\,
u\vert_{\partial\Omega}=f, \quad \mbox{supp}\,
f\subset\widetilde\Gamma, \,\,f\in H^\frac 32(\partial\Omega)\}.
$$
Here $\widetilde \Gamma$ is an arbitrarily fixed open subset of
$\partial\Omega.$ We set
$\Gamma_0=\partial\Omega\setminus\overline{\widetilde \Gamma}.$

In this paper, we consider the following inverse problem: {\it
Suppose that the partial Cauchy data $\mathcal C_{\lambda,\mu}$ are
given.  Can we determine the Lam\'e coefficients $\lambda$ and
$\mu$?}

This inverse problem has been studied since the 90's.
A linearized version of this inverse
problem for full data was studied in \cite{I}.
In two dimensions,
 Akamatsu, Nakamura and Steinberg \cite{A-N-S} proved that for the
case of full Cauchy data $(\widetilde \Gamma=\partial\Omega)$ one
can recover the Taylor series of the Lam\'e parameters on the
boundary provided that the Lam\'e coefficients are $C^\infty$
functions. This boundary determination result was extended in
\cite{NGb} to higher dimensions. In \cite{NG1} Nakamura and Uhlmann,
for the case of full Cauchy data, established that in the two
dimensions the Lam\'e coefficients are uniquely determined, assuming
that they are sufficiently close to a pair of positive constants.
Recently Imanuvilov and Yamamoto in \cite{IU2} proved for the two
dimensional case that the Lam\'e coefficient $\lambda$ can be
recovered from partial Cauchy data if the coefficient $\mu$ is some
positive constant. For the three dimensional case Nakamura and
Uhlmann in \cite{NG2}, \cite{NG3} and independently in \cite{E-R}
Eskin and Ralston proved uniqueness for both Lam\'e coefficients
provided that $\mu$ is close to a positive constant.  The proofs in
the above papers rely on construction of complex geometric optics
solutions. On the other hand, unlike the case of the Schr\"odinger
operator, for partial Cauchy data, the construction of such a
solutions for the isotropic Lam\'e system seems to be possible only
for a dense set of Lam\'e coefficients.

The recovery of Lam\'e coefficients by partial Cauchy data on
an arbitrary subboundary is useful from the practical point of
view, because one can limit input and measurement subsets of
$\partial\Omega$ as much as possible.
To the best of our knowledge, there are no results on the unique
recovery of the Lam\'e coefficients from the partial Cauchy data in
the three dimensional case.  The purpose of this article is to
prove such uniqueness in three dimensions.


Finally we mention that this inverse problem is closely related to
the method  known as Electrical Impedance Tomography (EIT). EIT
is used in prospection of oil and minerals
and in medical imaging in detecting breast cancer, pulmonary edema, etc.
For the mathematical treatments of this problem, we refer to \cite{AP},
\cite{Bu}, \cite{C}, \cite{IUY}, \cite{IUY1}, \cite{N}, \cite{SU3}
and the review paper \cite{U}.
\\
\vspace{0.3cm}

Our main result is the following theorem.
\begin{theorem}\label{main}
Let $\mu_1,\mu_2$ be some positive  constants and $\lambda_1,\lambda_2\in
 C^{\infty}(\overline\Omega)$ be some functions satisfying (\ref{Lame})
and $\lambda_1=\lambda_2$ on $\Gamma_0.$
 If $\mathcal C_{\lambda_1,\mu_1}=\mathcal C_{\lambda_2,\mu_2}$, then
 $(\lambda_1,\mu_1)=(\lambda_2,\mu_2).$
 \end{theorem}

{\bf Proof}. The proof consists in showing that from partial Cauchy
data one can recover the full Cauchy data. First following
\cite{NGb} we obtain that
\begin{equation}\label{01}
(\lambda_1,\mu_1)=(\lambda_2,\mu_2)\quad \mbox{on}
\,\,\widetilde \Gamma.
\end{equation}

Let $u_j\in H^2(\Omega)$ be a functions such that
\begin{equation}\label{00}
\mathcal L_{\lambda_j,\mu_j}(x,D)u_j=0\quad\mbox{in}\,\,\Omega,
\quad u_j\vert_{\partial \Omega}=f,
\end{equation}
where $\mbox{supp} f\subset \widetilde\Gamma.$ Since the partial Cauchy
data are the same, we obtain
\begin{equation}\label{000}
\Lambda_{\lambda_1,\mu_1}u_1=\Lambda_{\lambda_2,\mu_2}u_2\quad
\mbox{on}\,\,\widetilde \Gamma.
\end{equation}
 This equality combined with (\ref{01}) implies that
\begin{equation}\label{02}
(u_1,\frac{\partial u_1}{\partial \nu})=(u_2,\frac{\partial
u_2}{\partial \nu}) \quad \mbox{on}\,\,\widetilde \Gamma .
\end{equation}

Since the functions $\mu_j$ are assumed to be constants, from
(\ref{01}) we conclude that
\begin{equation}\label{03}
\mu_1=\mu_2\quad \mbox{on}\,\,\Omega.
\end{equation}
For constant $\mu$, we note that
$$
\mathcal L_{\lambda,\mu}(x,D)u = \mu\Delta u + (\mu+\lambda)\nabla
\mbox{div} \thinspace u + (\mbox{div}\thinspace u)\nabla\lambda.
$$
Applying to equation (\ref{00}) the operator $\mbox{rot}$ and using
the fact that $\mu_j$ is constant, we obtain
\begin{equation}\label{04}
\mu_j\Delta \mbox{rot} u_j=0\quad\mbox{in}\,\,\Omega.
\end{equation}

From (\ref{01}), (\ref{02}) and  equation (\ref{00}) we conclude
\begin{equation}\label{05}
(u_1,\frac{\partial u_1}{\partial \nu},
\partial_{x_ix_k} u_1)=(u_2,\frac{\partial u_2}{\partial \nu},
\partial_{x_ix_k} u_2) \quad
\mbox{on}\,\,\widetilde \Gamma, \quad \forall i,k\in \{1,2,3\}.
\end{equation}

Hence
\begin{equation}\label{06}
(\mbox{rot}\,u_1,\frac{\partial \mbox{rot}\,u_1}{\partial
\nu})=(\mbox{rot}\,u_2,\frac{\partial \mbox{rot}\,u_2}{\partial
\nu}) \quad \mbox{on}\,\,\widetilde \Gamma .
\end{equation}

Equality (\ref{06}) and the uniqueness of the solution for the Cauchy
problem for the Laplace equation imply
\begin{equation}\label{07}
\mbox{rot}\,u_1=\mbox{rot}\,u_2\quad \mbox{in}\,\,\Omega.
\end{equation}

The Lam\'e operator, with the coefficient $\mu=const,$ can be
written in the form $\mathcal L(x,D)u=\nabla
((\lambda+2\mu)\mbox{div}\,u) - \mu\mbox{rot rot} u. $ Then  using
(\ref{03}), (\ref{07}) we obtain

\begin{equation}\label{08}
\nabla ((\lambda_1+2\mu_1)\mbox{div}\,u_1)=\nabla
((\lambda_2+2\mu_1)\mbox{div}\,u_2)\quad \mbox{in}\,\,\Omega.
\end{equation}

Since
$(\lambda_1+2\mu_1)\mbox{div}\,u_1=(\lambda_2+2\mu_1)\mbox{div}\,u_2$
on $\widetilde \Gamma$, equation (\ref{08}) implies
\begin{equation}\label{09}
(\lambda_1+2\mu_1)\mbox{div}\,u_1=(\lambda_2+2\mu_1)\mbox{div}\,u_2\quad
\mbox{in}\,\,\Omega.
\end{equation}
From (\ref{03}), (\ref{09}), (\ref{07}) and the assumption
$(\lambda_1-\lambda_2)\vert_{\Gamma_0}=0$ we conclude
\begin{equation}\label{10}
\frac{\partial u_1}{\partial \nu}=\frac{\partial u_2}{\partial
\nu}\quad \mbox{on $\Gamma_0$}.
\end{equation}
Therefore if $\mbox{supp} \,f \subset \widetilde{\Gamma}$ in
(\ref{01})and $f\in H^\frac 32(\partial\Omega)$ then $\frac{\partial
u_1}{\partial\nu}=\frac{\partial u_2}{\partial\nu}$ on
$\partial\Omega$.

Next let $f\in H^\frac 32(\partial\Omega)$ and the functions $v_j\in
H^2(\Omega)$ be solutions of the following boundary value problem

\begin{equation}\label{11}
\mathcal L_{\lambda_j,\mu_j}(x,D) v_j=0\quad\mbox{in}\,\Omega,\quad
v_j\vert_{\partial\Omega}=f, \quad j\in \{1,2\}.
\end{equation}

We claim that
\begin{equation}\label{12}
\frac{\partial v_1}{\partial\nu}=\frac{\partial
v_1}{\partial\nu}\quad \mbox{on}\,\,\widetilde \Gamma .
\end{equation}

Indeed, let $w_j\in H^2(\Omega)$ be  a solution to the Lam\'e system
\begin{equation}\label{13}
\mathcal L_{\lambda_j,\mu_j}(x,D) w_j=0\quad\mbox{in $\Omega$},\quad
v_j\vert_{\partial\Omega}=g, \quad j\in \{1,2\},
\end{equation}
where $g\in H^\frac 32(\partial\Omega)$ and $\mbox{supp}\, g\subset
\widetilde \Gamma$ is an arbitrary function. Taking the scalar product of
equation (\ref{11}) with $w_j$ and integrating by parts, we have
$$
0=\int_\Omega( \mathcal L_{\lambda_j,\mu_j}(x,D) v_j,w_j)dx=
\int_\Omega(  v_j, \mathcal
L_{\lambda_j,\mu_j}(x,D)w_j)dx+\int_{\partial\Omega}((\Lambda_{\lambda,\mu}v_j,
w_j)-(\Lambda_{\lambda,\mu}w_j, v_j))d\sigma
$$
$$
=\int_{\partial\Omega}((\Lambda_{\lambda_j,\mu_j}v_j,
g)-(\Lambda_{\lambda_j,\mu_j}w_j, f))d\sigma
=\int_{\widetilde\Gamma}(\Lambda_{\lambda_j,\mu_j}v_j,
g)d\sigma-\int_{\partial\Omega}(\Lambda_{\lambda_j,\mu_j}w_j,
f)d\sigma
$$
$$
=\int_{\widetilde\Gamma}(\Lambda_{\lambda_j,\mu_j}v_j,
g)d\sigma-\int_{\partial\Omega}(\Lambda_{\lambda_1,\mu_1}w_1,
f)d\sigma,
$$
where $d\sigma$ denotes the surface measure.

This integral identity implies
$$
\Lambda_{\lambda_1,\mu_1}v_1=\Lambda_{\lambda_2,\mu_2}v_2
\quad\mbox{on $\widetilde\Gamma$}.
$$
Repeating the arguments (\ref{04})-(\ref{10}) we conclude
\begin{equation}\label{popo}
\frac{\partial v_1}{\partial\nu}=\frac{\partial
v_2}{\partial\nu}\quad\mbox{on}\,\Gamma_0.
\end{equation}

Hence, by (\ref{10}), (\ref{popo}) the following full Cauchy data
are equal:
$$
\widetilde{\mathcal C}_{\lambda_1,\mu_1}=\widetilde{\mathcal
C}_{\lambda_2,\mu_2}
$$
where
$$
\widetilde{\mathcal C}_{\lambda,\mu}=\{ (u,\Lambda_{\lambda,\mu}
u)\vert_{\partial\Omega}; \thinspace \mathcal L_{\lambda,\mu}(x,D)
u=0\quad \mbox{in}\,\,\Omega,\,\, u\vert_{\partial\Omega}=f,
\,\,f\in H^\frac 32(\partial\Omega)\}.
$$

Applying the result of \cite{E-R}, \cite{NG2} and \cite{NG3}, we obtain
that $\lambda_1=\lambda_2.$ $\square$

This result immediately implies a local result for $\mu$ near constant.

\begin{corollary}Let $B$ be a bounded set in $C^\infty(\overline
\Omega)$, $\lambda_1,\lambda_2\in B , \lambda_1=\lambda_2$ on
$\Gamma_0$  and $\lambda_j(x)>C>0,\mu_j(x)>C>0 $ on
$\overline\Omega.$ There exist positive $\epsilon(B)>0$  and
positive sufficiently large number $N$ such that if
$\sum_{k=1}^2\Vert \nabla \mu_k\Vert_{C^N(\overline\Omega)}\le
\epsilon(B)$ and $\mathcal C_{\lambda_1,\mu_1}=\mathcal
C_{\lambda_2,\mu_2}$, then $(\lambda_1,\mu_1)=(\lambda_2,\mu_2).$
\end{corollary}

{\bf Proof.} Our proof  by contradiction. Suppose that the statement
of the corollary is false. Then there exists a sequence of positive
$\{\epsilon_j\}_{j=1}^\infty$ such that $\epsilon_j\rightarrow 0$
and for each $\epsilon_j$ there exists
$\{(\lambda_{k,j},\mu_{k,j})\}$ such that
\begin{equation}\label{15}
\{\lambda_{k,j}\}_{j=1}^\infty\subset
B,\,\,k=\{1,2\},\quad\mbox{and}\quad \sum_{k=1}^2\Vert \nabla
\mu_{k,j}\Vert_{C^N(\overline\Omega)}\le \epsilon_j.
\end{equation}
and \begin{equation}\label{16}\mathcal
C_{\lambda_{1,j},\mu_{1,j}}=\mathcal
C_{\lambda_{2,j},\mu_{2,j}}\quad \forall j\in \{1,\dots, \infty\}.
\end{equation}
By (\ref{15}) and (\ref{16}) there exist $\lambda_1,\lambda_2\in
C^\infty(\overline \Omega)$ and positive constants $\mu_k$ such that
$$
\mathcal C_{\lambda_{1},\mu_{1}}=\mathcal C_{\lambda_{2},\mu_{2}}.
$$
Applying Theorem \ref{main}, we complete the proof of the
corollary. $\square$

\end{document}